\documentclass[a4paper,12pt]{article}
\usepackage{geometry}
\usepackage{natbib}
\usepackage{amsmath}
\usepackage{amssymb}
\usepackage[utf8]{inputenc}
\usepackage{longtable}
\usepackage{lscape}
\bibpunct{(}{)}{;}{a}{,}{,} 
\usepackage[hyphens]{url}
\usepackage{breakurl} 
\usepackage{float}
\usepackage[font={sf,bf,small},textfont=md,textfont=md]{caption} 
\usepackage{setspace}

\usepackage{graphicx}
\usepackage{subfigure}
\usepackage{rotating}
\usepackage{booktabs}

 \renewcommand\appendix{\par 
    \setcounter{section}{0}%
    \setcounter{subsection}{0}%
    \setcounter{table}{0}%
    \setcounter{figure}{0}%
    \renewcommand\thesection{\Alph{section}}%
    \renewcommand*{\thefigure}{\Alph{section}S\arabic{figure}}%
    \renewcommand*{\thetable}{\Alph{section}S\arabic{table}}
     \renewcommand{\theequation}{A-\arabic{equation}} }
   
\title{Fluctuating selection models and McDonald-Kreitman type analyses}

\author{\begin{normalsize}\textbf{Note}\end{normalsize}
\\Toni I. Gossmann$^{1,2}$, David Waxman$^{3}$, and Adam Eyre-Walker$^{1,\ast}$\\
\\
$^{1}$ School of Life Sciences, University of Sussex,\\ Brighton, United Kingdom
\\
$^{2}$ Department of Animal and Plant Sciences, University of Sheffield,\\ Sheffield, United Kingdom\\
$^{3}$ Centre for Computational Systems Biology,\\ Fudan University, Shanghai, China\\
\begin{small}$^{\ast}$Correspondence: a.c.eyre-walker@sussex.ac.uk\end{small}}

\begin{document}

\maketitle

\begin{abstract}
It is likely that the strength of selection acting upon a mutation varies through time due to changes in the environment. However, most population genetic theory assumes that the strength of selection remains constant. Here we investigate the consequences of fluctuating selection pressures on the quantification of adaptive evolution using McDonald-Kreitman (MK) style approaches. In agreement with previous work, we show that fluctuating selection can generate evidence of adaptive evolution even when the expected strength of selection on a mutation is zero. However, we also find that the mutations, which contribute to both polymorphism and divergence tend, on average, to be positively selected during their lifetime, under fluctuating selection models. This is because mutations that fluctuate, by chance, to positive selected values, tend to reach higher frequencies in the population than those that fluctuate towards negative values. Hence the evidence of positive adaptive evolution detected under a fluctuating selection model by MK type approaches is genuine since fixed mutations tend to be advantageous on average during their lifetime. Never-the-less we show that methods tend to underestimate the rate of adaptive evolution when selection fluctuates. 
\end{abstract}

\clearpage

The McDonald-Kreitman (MK) test \citep{McDonald1991}, and its derivatives \citep{Fay2001, Smith2002, Bierne2004, Eyre-Walker2009} use the contrast between the levels of polymorphism and substitution at neutral and selected sites to infer the presence of adaptive evolution in the divergence between species. Modified versions of the MK test allow one to quantify $\alpha$, the proportion of nonsynonymous differences between species due to adaptive evolution \citep{Fay2001, Smith2002, Bierne2004, Eyre-Walker2009}. The MK test has been widely applied to a number of species and estimates of $\alpha$ vary substantially from limited evidence ($\alpha \approx 0$ to 10\%) in humans \citep{Sequencing2005, Zhang2005, Boyko2008a} and many plant species \citep{Gossmann2010} to more than 50\% in Drosophila \citep{Smith2002, Charlesworth2006}, some plants with large effective population size \citep{Slotte2010, Strasburg2011} and bacteria \citep{Charlesworth2006}.

The MK test framework implicitly assumes that selection pressures are constant. However, the environment for most organisms is constantly changing due to fluctuations in physical factors, such as temperature, and biotic factors, such as the prevalence of competitor species and the density and genotype frequencies of other conspecific individuals. This is likely to lead to changes in the strength of selection acting upon a mutation through time \citep{Bell2010}; in the extreme this might mean that a mutation is advantageous at one time-point, but deleterious at another. Despite the likelihood that selection fluctuates through time there is relatively little evidence that this is the case. This is probably because measuring the strength of selection is difficult and detecting fluctuating selection requires analyses over several years. However, analyses of data from several species have suggested that some polymorphisms are subject to fluctuating selection \citep[][reviewed by \citealp{Bell2010}]{Fisher1947,Mueller1985,Lynch1987,O'Hara2005}. In these examples there are changes in the frequency of mutations that appear to be too great to be explained by either random genetic drift or migration. In most of these analyses the mean strength of selection acting upon a mutation appears to be close to zero. However, this might be a sampling artifact, a mutation subject to fluctuating selection in which the average selection coefficient is non-zero is more likely to be lost or fixed.

Fluctuating selection is likely to be more prevalent than the few well documented examples suggest and \cite{Bell2010} has argued that fluctuating selection might help resolve why most traits show substantial heritability, even though selection on a short time-scale often appears to be quite strong. Despite the likelihood that the strength of selection varies most work in theoretical population genetics has assumed that the strength of selection is constant through time. Exceptions are the work by \cite{Kimura1954}, \cite{Gillespie1973, gillespie1991causes}, \cite{Jensen1973}, \cite{Karlin1974}, \cite{Takahata1975}, \cite{Huerta-Sanchez2008} and \cite{Waxman2011}. \cite{Huerta-Sanchez2008} have investigated how fluctuating selection affects the allele frequency distribution, and hence the site frequency spectrum (SFS), and the probability of fixation. They showed that although the expected strength of selection is zero, fluctuating selections leads to an increase in the probability of fixation, a decrease in diversity and a change in the SFS. \cite{Huerta-Sanchez2008} therefore suggested that fluctuating selection might generate artifactual evidence of adaptive evolution artifactutal evidence of adaptive evolution in MK type analyses. Here we investigate whether this indeed the case and analyse the average strength of selection of mutations contributing to divergence and polymorphism.

We use the basic two-allele model investigated by \cite{Huerta-Sanchez2008} in which the strength of selection varies each generation, with the expected strength of selection acting upon each allele being zero. \cite{Huerta-Sanchez2008} show that this model may be summarized in terms of the parameter $\beta = 2N_e \mathbf{Var}(s)$ where $N_e$ is the effective population size and $\mathbf{Var}(s)$ is the variance in the strength of selection. Furthermore, they demonstrate that this model behaves identically in terms of the SFS and probability of fixation to one in which the strength of selection is autocorrelated between generations; the autocorrelation simply increases the value of $\beta$. We investigated the model of \cite{Huerta-Sanchez2008} by simulation so that we can track the strength of selection acting upon each mutation as it segregates in the population. In our haploid simulation we introduce a new mutation at a frequency of $1/N$, where $N$ is the population size, at a site that is monomorphic. The strength of selection acting upon the two alleles is then drawn from a normal distribution with a mean of 1 and a standard deviation of $\sigma$. Using the frequency of the new mutation and the strength of selection acting upon the two alleles, we calculated the expected frequency of the new mutation in the next generation, $\hat{f}$, and generated the actual frequency, $f$, as a number drawn from a binomial distribution with sample size $N$ and  probability $\hat{f}$. If the mutation is lost or fixed a new mutation is introduced and the simulation repeated, otherwise new selection strengths are sampled and another generation repeated. For each value of $\beta$ we simulated the evolution of 100,000 mutations. We used the simulated data to infer the expected SFS for a population sample of 20 chromosomes (similar results were obtained for other sample sizes).

For each mutation that is fixed, or that reaches any arbitrary frequency $f$, we can calculate the mean strength of selection that has acted upon that mutation up to the time that we sample it. Let us define the true value of $\alpha$ as the proportion of substitutions in which all mutations that have fixed have a mean strength of selection that is positive, during their passage through the population.
\begin{align}
\alpha_{True} &= \frac{D_{n_{adaptive}}}{D_n}
\end{align}
where $D_n$ is the number of nonsynonymous substitutions and $D_{n_{adaptive}}$ is the number of nonsynonymous substitutions with a mean positive strength of selection at the time of fixation. We estimated $\alpha$ using several commonly used methods. First we applied the method of \cite{Fay2001}:
\begin{align}
\alpha &= 1 - \frac{D_sP_n}{D_nP_s}
\end{align}
where $D_n$, $D_s$, $P_n$ and $P_s$ are the numbers of nonsynonymous and synonymous substitutions and polymorphisms, respectively. This method does not take into account the effect of slightly deleterious mutations, which tend to bias the estimate of $\alpha$ downwards. We therefore applied two methods that attempt to correct for this bias. The first is the method of \cite{Eyre-Walker2009}, and second the method of \cite{Schneider2011}. The method of \cite{Eyre-Walker2009} assumes that advantageous mutations are strongly selected and do not contribute substantially to polymorphism. The method of \cite{Schneider2011} does not make this assumption and attempts to infer the proportion of mutations that are advantageous and the strength of selection acting in favour of them. For each method we contrast what happens at sites subject to a certain level of fluctuating selection to those at which there is no fluctuation. In both cases the expected strength of selection is zero; the sites with no fluctuation are therefore evolving neutrally. The simulation is set up such that there is free recombination between sites.

\begin{figure}
\center
\subfigure[]{{\includegraphics[width=0.3\textwidth]{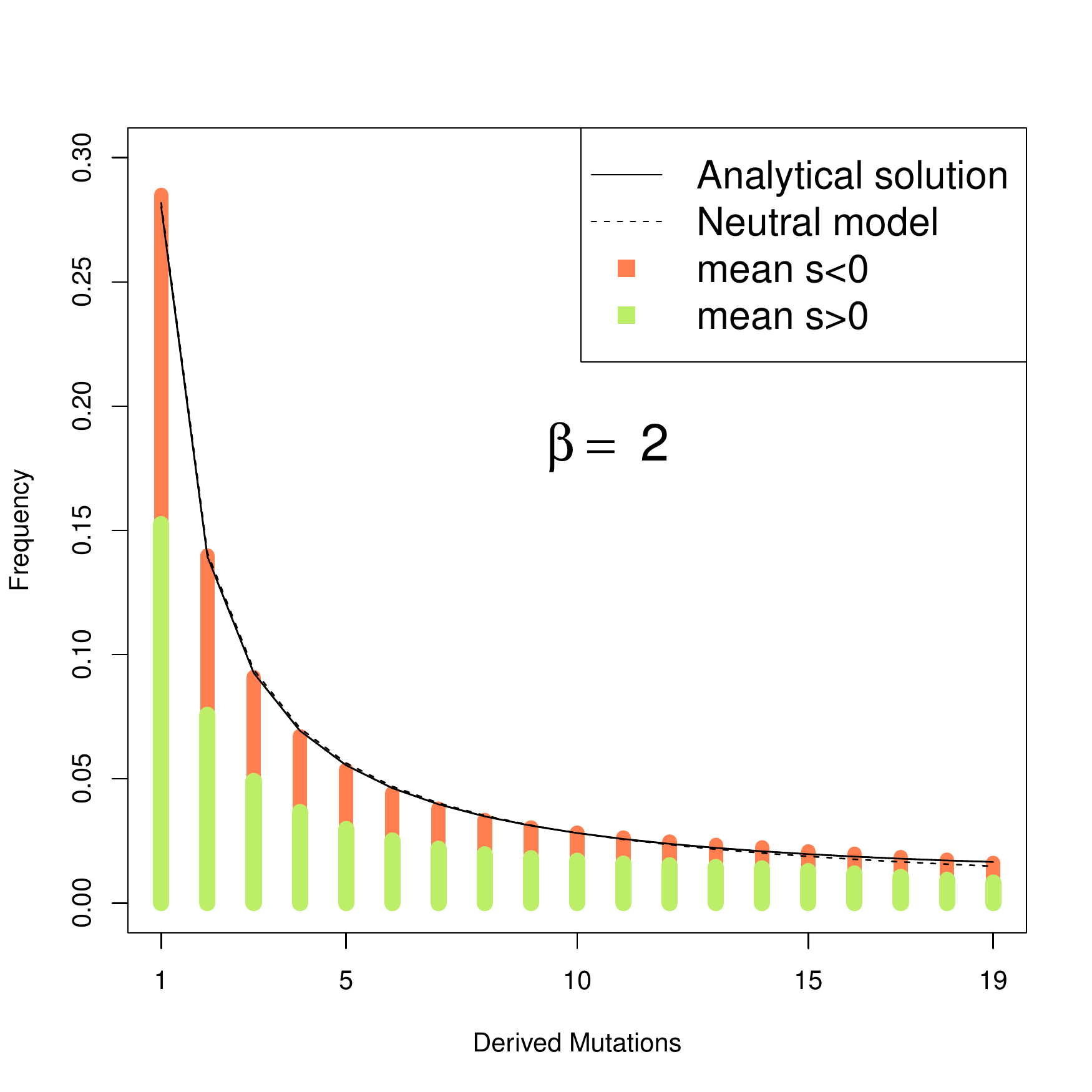}}}
\subfigure[]{{\includegraphics[width=0.3\textwidth]{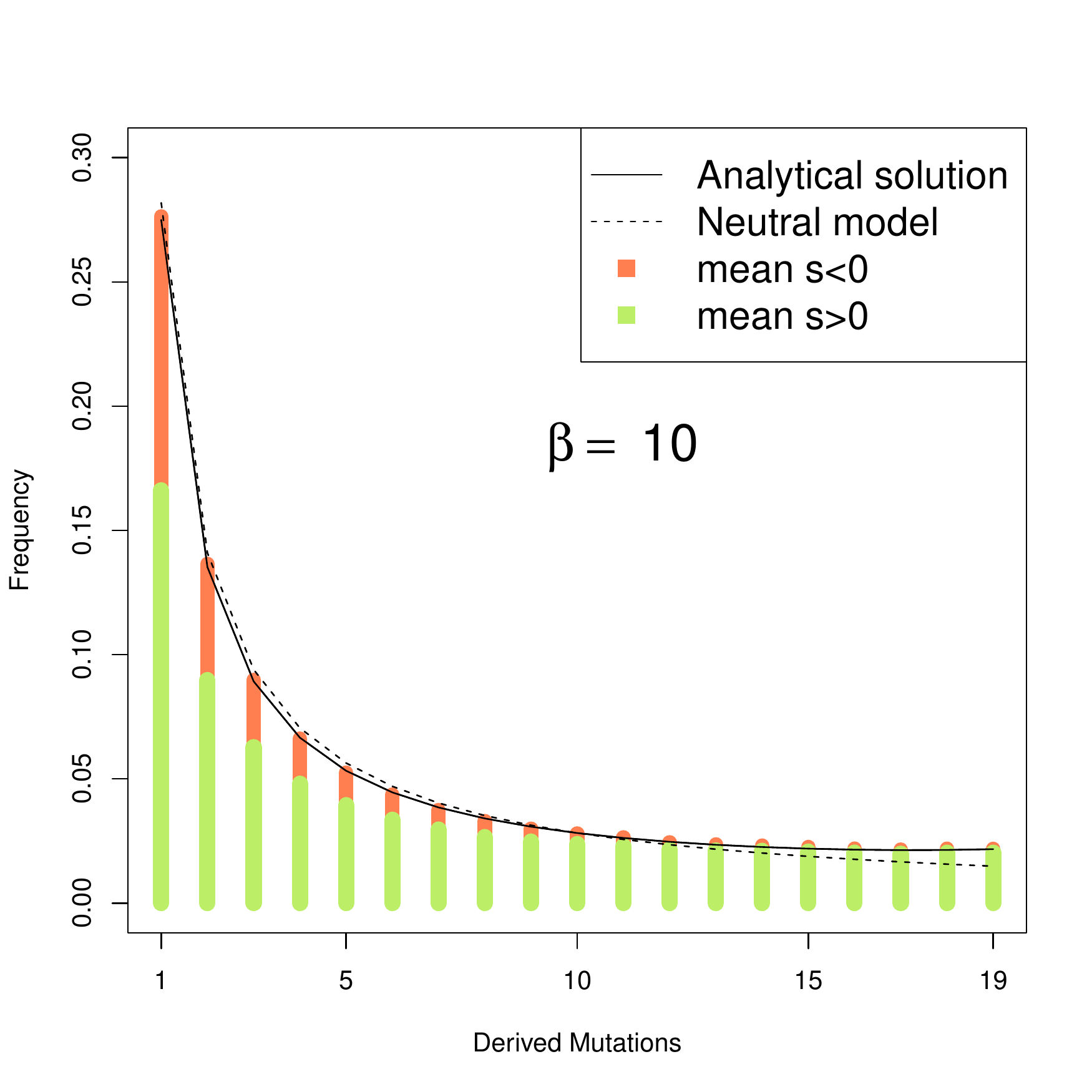}}}
\subfigure[]{{\includegraphics[width=0.3\textwidth]{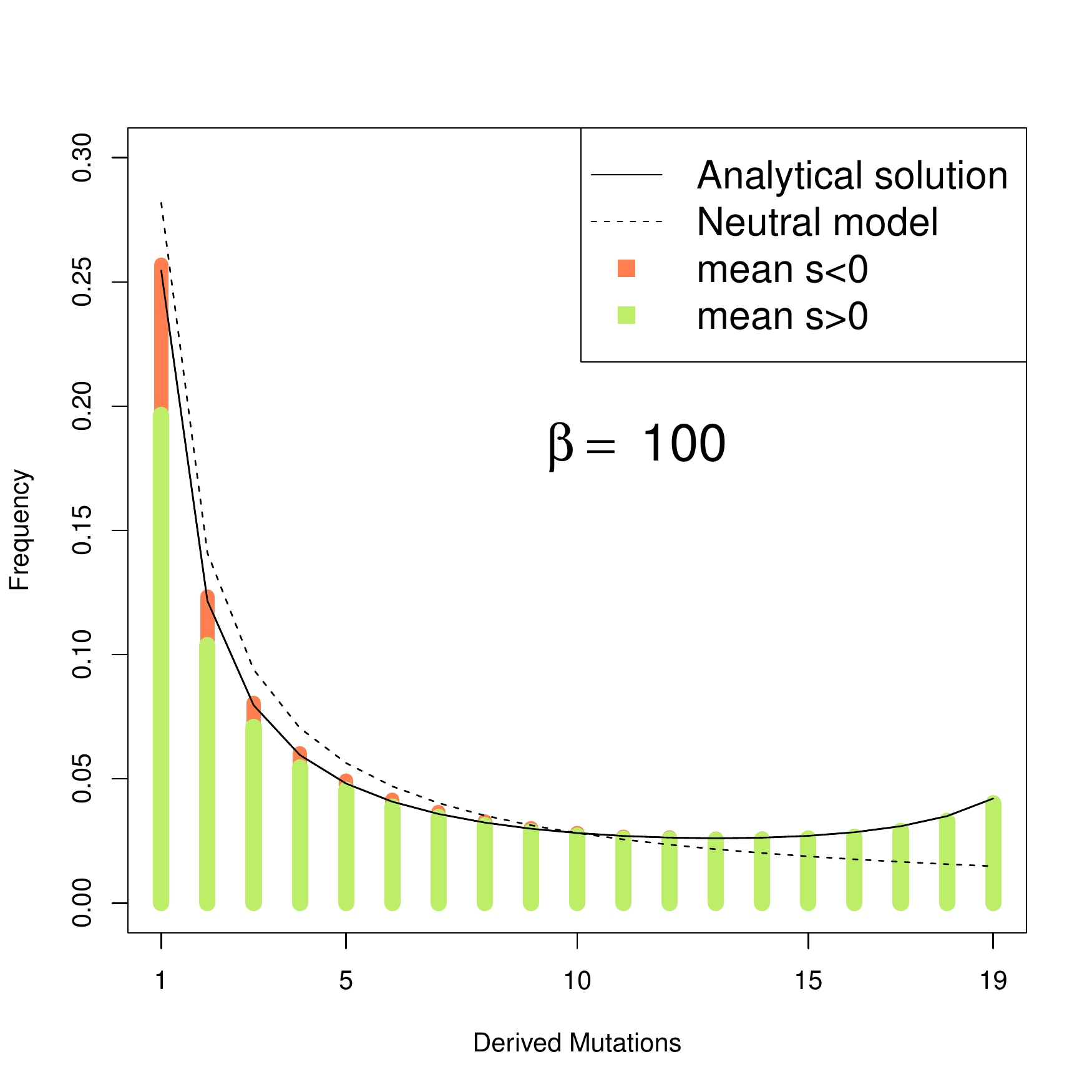}}}
 \caption{SFS generated under fluctuating conditions ($\beta=2$, $10$ and $100$) with mean selective effect of zero. The proportion of mutations with positive and negative mean selection coefficients are shown in green and red, respectively. The analytical solution is obtained from the equations from \cite{Huerta-Sanchez2008}.}{}
 \label{SFS}
\end{figure}

\begin{figure}
\center
\subfigure[]{{\includegraphics[angle=270,width=0.3\textwidth]{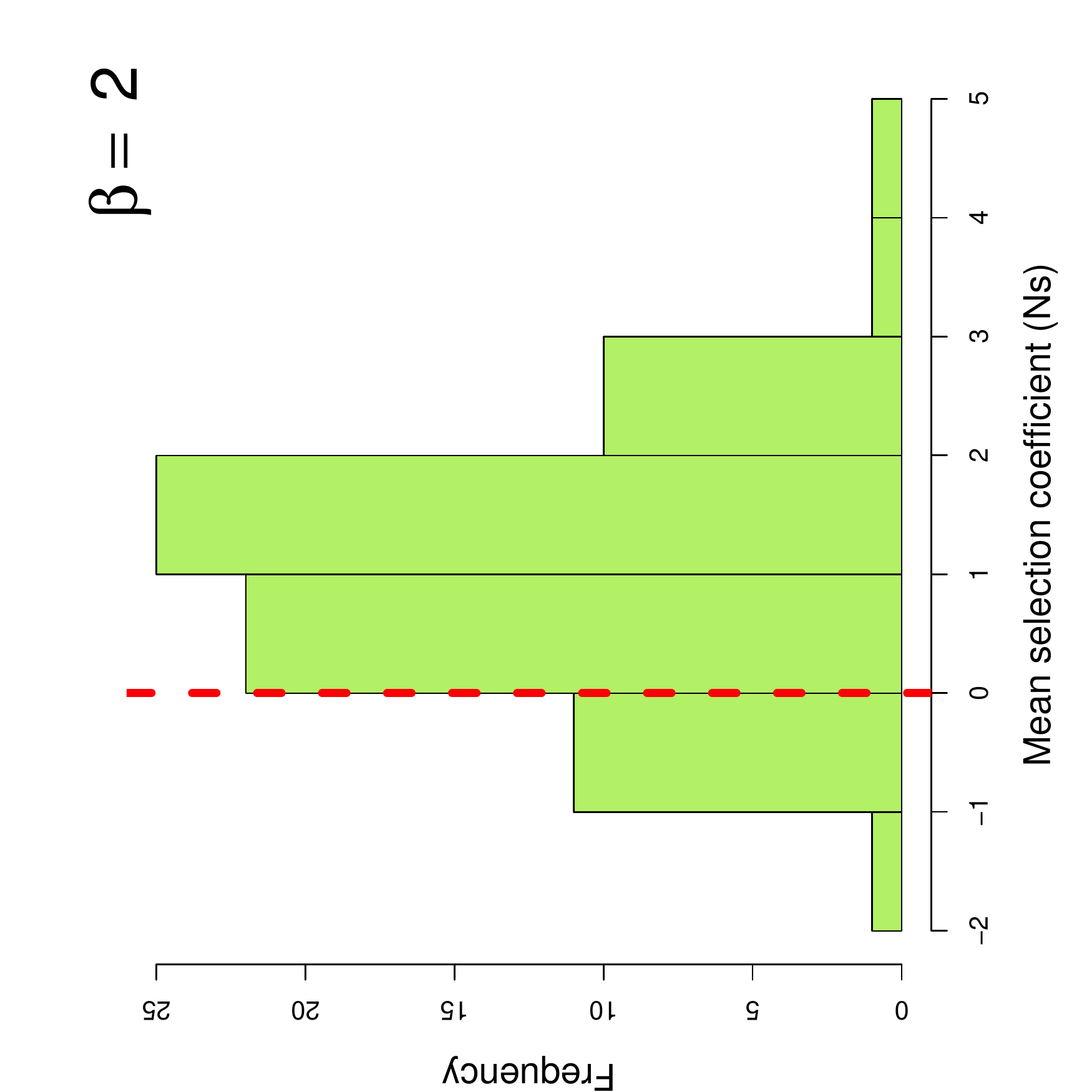}}}
\subfigure[]{{\includegraphics[angle=270,width=0.3\textwidth]{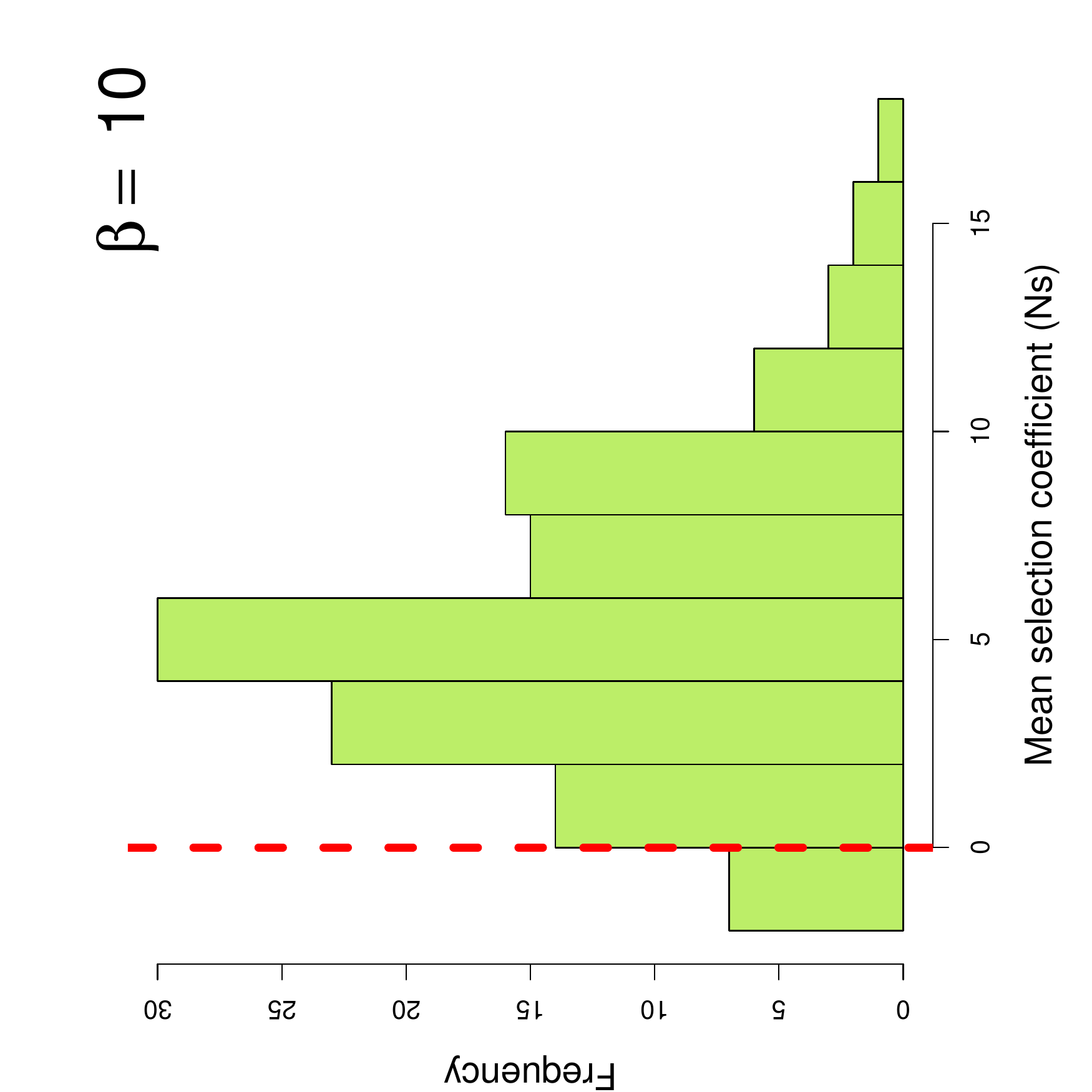}}}
\subfigure[]{{\includegraphics[angle=270,width=0.3\textwidth]{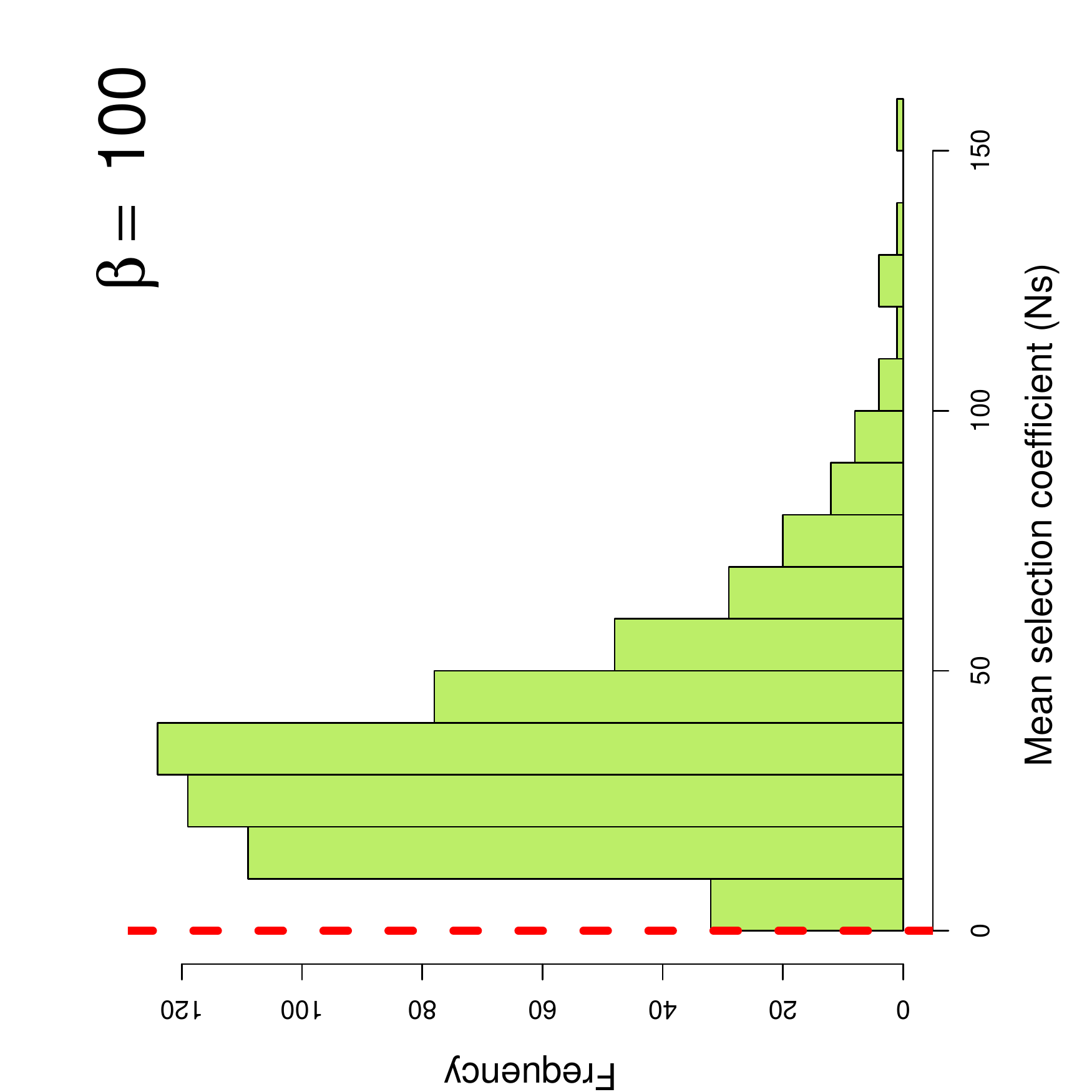}}}
 \caption{Distributions of mean fitness effects of mutations at the time of fixation for fluctuating conditions ($\beta=2$, $10$ and $100$) with mean selective effect for all mutations of zero.}
 \label{fixed}
\end{figure}

Our simulations demonstrate that fluctuating selection can generate evidence of adaptive evolution; all three of the methods to estimate $\alpha$ yield positive estimates for all values of $\beta$ (Table \ref{mk}). The fact that a fluctuating selection model generates evidence of adaptive evolution even when the expected strength of selection is zero suggests that fluctuating selection generates artifactual evidence of positive selection \citep{Huerta-Sanchez2008}. However, the mean strength of selection experienced by the mutation, that is sampled in a set of DNA sequences, or that spreads to fixation, might not be zero, even though its expected value over all mutations (not just those that fix) is zero; it might be that those mutations which spread to high frequency in the population are those, which just by chance have mean selective values that are positive, whilst those mutations which fluctuate to negative values are lost from the population. To investigate this we tracked the mean strength of selection of each mutation at each frequency up to when it was lost or fixed. From this analysis it is evident that the vast majority of mutations that contribute to the SFS are positively selected, except at very low frequencies and when fluctuations in the strength of selection are quite weak (Figure \ref{SFS}). The bias towards positive mean strengths of selection is even more extreme for those mutations that become fixed (Figure \ref{fixed}).

\begin{figure}
\center
\subfigure[]{{\includegraphics[angle=270,width=0.3\textwidth]{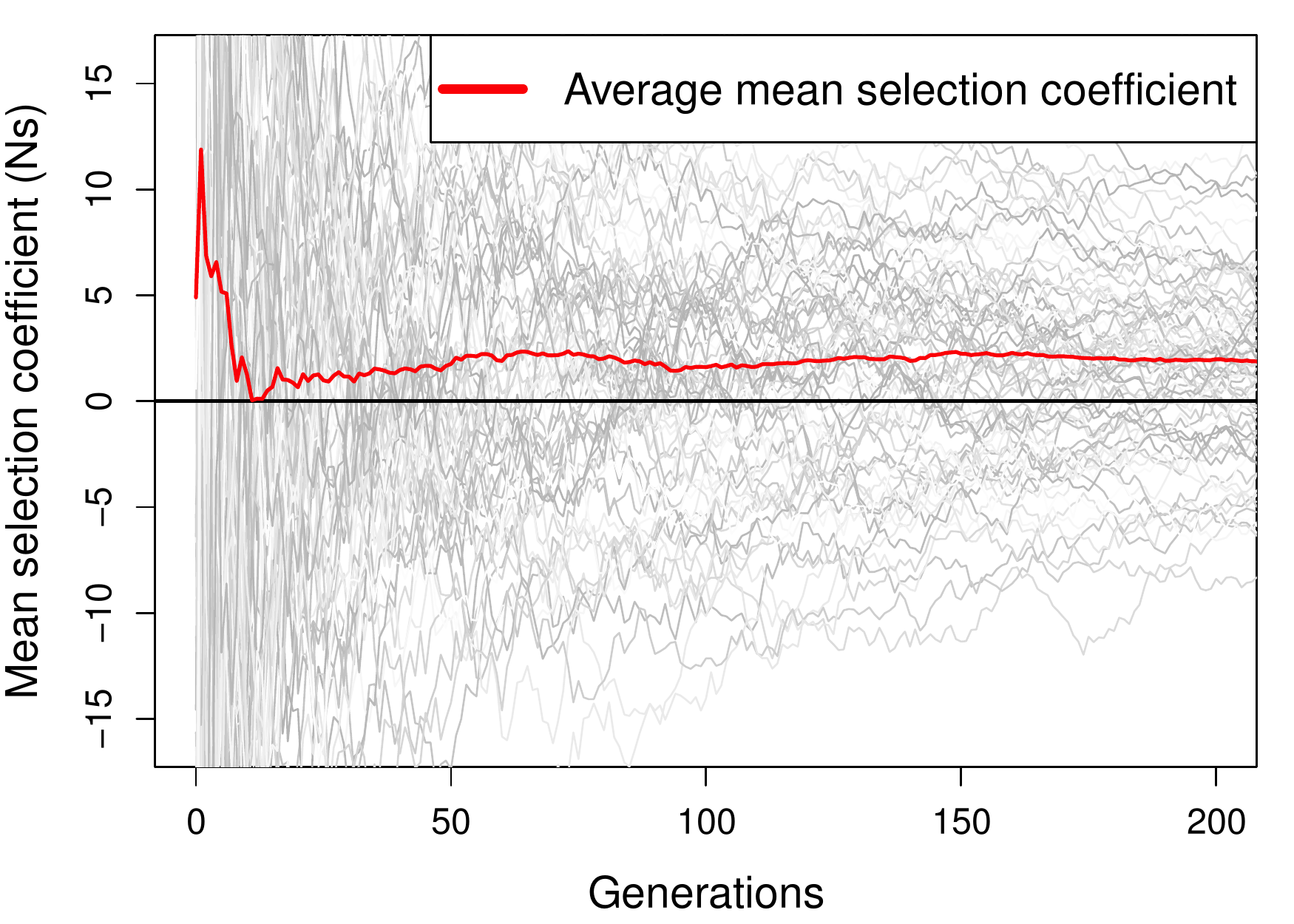}}}
\subfigure[]{{\includegraphics[angle=270,width=0.3\textwidth]{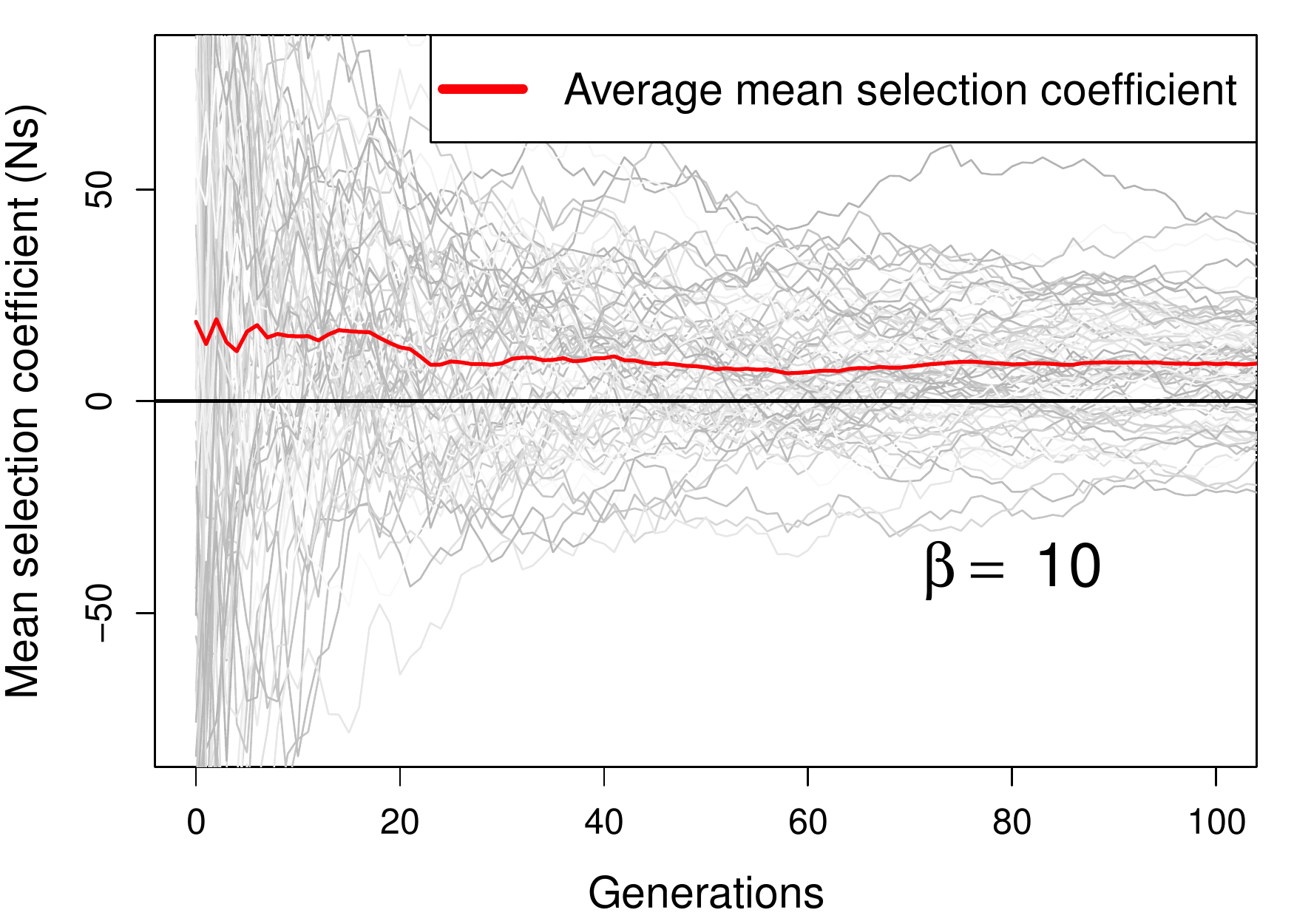}}}
\subfigure[]{{\includegraphics[angle=270,width=0.3\textwidth]{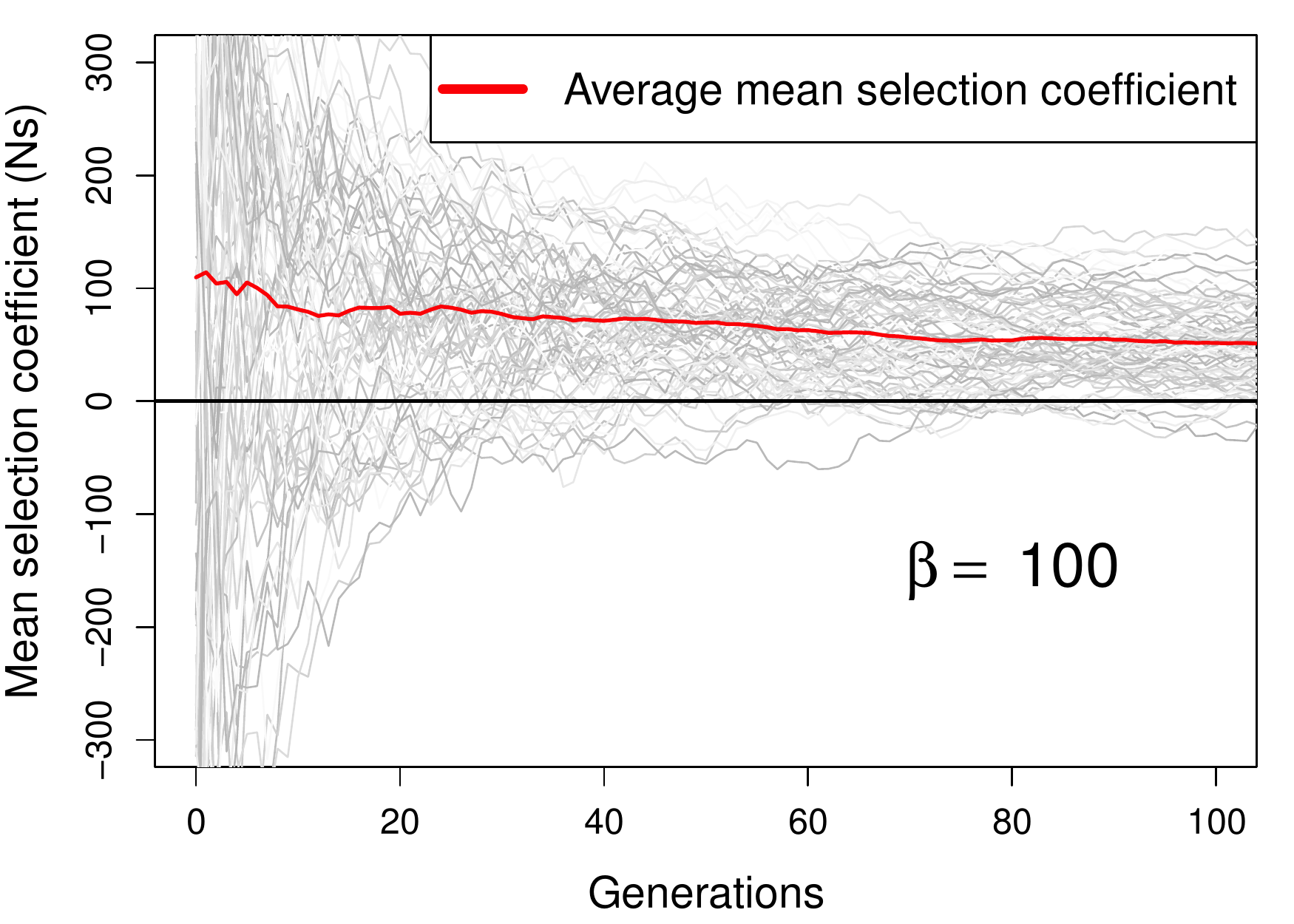}}}
\caption{
{\bf Average mean selection coefficient over time for fixed mutations under fluctuating conditions of $\beta=2, 10, 100$.} Shown are the first 100 generations (200 in case of $\beta=2$) of 80 mutations that got fixed. The red line indicates the average mean selection coefficient, trajectories in grayscale indicate mean selection coefficients for individual mutations.}
\label{fixed_time}
\end{figure}

If we track mutations that ultimately become fixed it is evident that those mutations that start off being slightly negative quickly become positive in their mean value (Figure \ref{fixed_time}). Interestingly those that start off being highly positive tend to decrease in mean selection coefficient as well; this is probably a consequence of averaging over many selective episodes, and hence approaching the expected value. We also find that the average mean selection coefficient for all mutations that get fixed declines with time. This is because the critical time for an advantageous mutation is when it is rare because it is more likely to be lost. Those mutations that are strongly positively selected at an early stage have more chance of remaining in the population. 

From our simulations it is possible to obtain $\alpha_{True}$, the proportion of fixations that have a mean selection coefficient that is positive and can therefore be considered as truly adaptive. The methods of \cite{Fay2001}, \cite{Eyre-Walker2009} and \cite{Schneider2011} all consistently underestimate $\alpha_{True}$, although the effect appears to be more severe when fluctuating conditions are weak. The method of  \cite{Schneider2011} is better than that of \cite{Eyre-Walker2009}, which is better than that of \cite{Fay2001}. It is perhaps not surprising that the method of \cite{Schneider2011} performs best since the vast majority of mutations that become fixed have a small average positive selection coefficient and such mutations are likely to contribute to polymorphism, which the other two methods assume does not happen.

We find, in agreement with the suggestion of \cite{Huerta-Sanchez2008}, that the fluctuating selection does lead to a signature of adaptive evolution. However, we also show that those mutations contributing to polymorphism and divergence are on average positively selected during their lives, even though the expected strength selection is zero. We therefore conclude that the signature of adaptive evolution is genuine. However, it is also evident that methods to estimate the level of adaptive evolution tend to underestimate the contribution of mutations subject to fluctuating selection.

\renewcommand\refname{Literature Cited} 
 \bibliographystyle{genetics}  

\begin{thebibliography}{26}
\expandafter\ifx\csname natexlab\endcsname\relax\def\natexlab#1{#1}\fi

\bibitem[{{\sc Bell}(2010)}]{Bell2010}
{\sc Bell, G.}, 2010 {F}luctuating selection: the perpetual renewal of
  adaptation in variable environments.
\newblock Philos Trans R Soc Lond B Biol Sci {\bf 365}: 87--97.

\bibitem[{{\sc Bierne} and {\sc Eyre-Walker}(2004)}]{Bierne2004}
{\sc Bierne, N.}, and {\sc A.~Eyre-Walker}, 2004 {T}he genomic rate of adaptive
  amino acid substitution in {D}rosophila.
\newblock Mol Biol Evol {\bf 21}: 1350--1360.

\bibitem[{{\sc Boyko} {\em et~al.\/}(2008){\sc Boyko}, {\sc Williamson}, {\sc
  Indap}, {\sc Degenhardt}, {\sc Hernandez} {\em et~al.\/}}]{Boyko2008a}
{\sc Boyko, A.~R.}, {\sc S.~H. Williamson}, {\sc A.~R. Indap}, {\sc J.~D.
  Degenhardt}, {\sc R.~D. Hernandez}, {\em et~al.\/}, 2008 {A}ssessing the
  evolutionary impact of amino acid mutations in the human genome.
\newblock PLoS Genet {\bf 4}: e1000083.

\bibitem[{{\sc Charlesworth} and {\sc Eyre-Walker}(2006)}]{Charlesworth2006}
{\sc Charlesworth, J.}, and {\sc A.~Eyre-Walker}, 2006 {T}he rate of adaptive
  evolution in enteric bacteria.
\newblock Mol Biol Evol {\bf 23}: 1348--1356.

\bibitem[{{\sc {Chimpanzee Sequencing and Analysis
  Consortium}}(2005)}]{Sequencing2005}
{\sc {Chimpanzee Sequencing and Analysis Consortium}}, 2005 {I}nitial sequence
  of the chimpanzee genome and comparison with the human genome.
\newblock Nature {\bf 437}: 69--87.

\bibitem[{{\sc Eyre-Walker} and {\sc Keightley}(2009)}]{Eyre-Walker2009}
{\sc Eyre-Walker, A.}, and {\sc P.~D. Keightley}, 2009 {E}stimating the rate of
  adaptive molecular evolution in the presence of slightly deleterious
  mutations and population size change.
\newblock Mol Biol Evol {\bf 26}: 2097--2108.

\bibitem[{{\sc Fay} {\em et~al.\/}(2001){\sc Fay}, {\sc Wyckoff} and {\sc
  Wu}}]{Fay2001}
{\sc Fay, J.~C.}, {\sc G.~J. Wyckoff}, and {\sc C.~I. Wu}, 2001 {P}ositive and
  negative selection on the human genome.
\newblock Genetics {\bf 158}: 1227--1234.

\bibitem[{{\sc Fisher} and {\sc Ford}(1947)}]{Fisher1947}
{\sc Fisher, R.}, and {\sc E.~Ford}, 1947 {T}he spread of a gene in natural
  conditions in a colony of the moth {P}anaxia dominula {L}.
\newblock Heredity {\bf 1}: 143--174.

\bibitem[{{\sc Gillespie}(1973)}]{Gillespie1973}
{\sc Gillespie, J.}, 1973 {N}atural selection with varying selection
  coefficients—a haploid model.
\newblock Genet. Res {\bf 21}: 115--120.

\bibitem[{{\sc Gillespie}(1991)}]{gillespie1991causes}
{\sc Gillespie, J.}, 1991 {\em {T}he causes of molecular evolution\/},
  volume~2.
\newblock Oxford University Press, USA.

\bibitem[{{\sc Gossmann} {\em et~al.\/}(2010){\sc Gossmann}, {\sc Song}, {\sc
  Windsor}, {\sc Mitchell-Olds}, {\sc Dixon} {\em et~al.\/}}]{Gossmann2010}
{\sc Gossmann, T.~I.}, {\sc B.-H. Song}, {\sc A.~J. Windsor}, {\sc
  T.~Mitchell-Olds}, {\sc C.~J. Dixon}, {\em et~al.\/}, 2010 {G}enome wide
  analyses reveal little evidence for adaptive evolution in many plant species.
\newblock Mol Biol Evol {\bf 27}: 1822--1832.

\bibitem[{{\sc Huerta-Sanchez} {\em et~al.\/}(2008){\sc Huerta-Sanchez}, {\sc
  Durrett} and {\sc Bustamante}}]{Huerta-Sanchez2008}
{\sc Huerta-Sanchez, E.}, {\sc R.~Durrett}, and {\sc C.~D. Bustamante}, 2008
  {P}opulation genetics of polymorphism and divergence under fluctuating
  selection.
\newblock Genetics {\bf 178}: 325--337.

\bibitem[{{\sc Jensen}(1973)}]{Jensen1973}
{\sc Jensen, L.}, 1973 {R}andom selective advantages of genes and their
  probabilities of fixation.
\newblock Genet Res {\bf 21}: 215--219.

\bibitem[{{\sc Karlin} and {\sc Levikson}(1974)}]{Karlin1974}
{\sc Karlin, S.}, and {\sc B.~Levikson}, 1974 {T}emporal fluctuations in
  selection intensities: {C}ase of small population size.
\newblock Theoretical Population Biology {\bf 6}: 383 -- 412.

\bibitem[{{\sc Kimura}(1954)}]{Kimura1954}
{\sc Kimura, M.}, 1954 {P}rocess {L}eading to {Q}uasi-{F}ixation of {G}enes in
  {N}atural {P}opulations {D}ue to {R}andom {F}luctuation of {S}election
  {I}ntensities.
\newblock Genetics {\bf 39}: 280--295.

\bibitem[{{\sc Lynch}(1987)}]{Lynch1987}
{\sc Lynch, M.}, 1987 {T}he consequences of fluctuating selection for isozyme
  polymorphisms in {D}aphnia.
\newblock Genetics {\bf 115}: 657--669.

\bibitem[{{\sc McDonald} and {\sc Kreitman}(1991)}]{McDonald1991}
{\sc McDonald, J.~H.}, and {\sc M.~Kreitman}, 1991 {A}daptive protein evolution
  at the {A}dh locus in {D}rosophila.
\newblock Nature {\bf 351}: 652--654.

\bibitem[{{\sc Mueller} {\em et~al.\/}(1985){\sc Mueller}, {\sc Barr} and {\sc
  Ayala}}]{Mueller1985}
{\sc Mueller, L.~D.}, {\sc L.~G. Barr}, and {\sc F.~J. Ayala}, 1985 {N}atural
  selection vs. random drift: evidence from temporal variation in allele
  frequencies in nature.
\newblock Genetics {\bf 111}: 517--554.

\bibitem[{{\sc O'Hara}(2005)}]{O'Hara2005}
{\sc O'Hara, R.~B.}, 2005 {C}omparing the effects of genetic drift and
  fluctuating selection on genotype frequency changes in the scarlet tiger
  moth.
\newblock Proc Biol Sci {\bf 272}: 211--217.

\bibitem[{{\sc Schneider} {\em et~al.\/}(2011){\sc Schneider}, {\sc
  Charlesworth}, {\sc Eyre-Walker} and {\sc Keightley}}]{Schneider2011}
{\sc Schneider, A.}, {\sc B.~Charlesworth}, {\sc A.~Eyre-Walker}, and {\sc
  P.~D. Keightley}, 2011 {A} {M}ethod for {I}nferring the {R}ate of
  {O}ccurrence and {F}itness {E}ffects of {A}dvantageous {M}utations.
\newblock Genetics .

\bibitem[{{\sc Slotte} {\em et~al.\/}(2010){\sc Slotte}, {\sc Foxe}, {\sc
  Hazzouri} and {\sc Wright}}]{Slotte2010}
{\sc Slotte, T.}, {\sc J.~P. Foxe}, {\sc K.~M. Hazzouri}, and {\sc S.~I.
  Wright}, 2010 {G}enome-wide evidence for efficient positive and purifying
  selection in {C}apsella grandiflora, a plant species with a large effective
  population size.
\newblock Mol Biol Evol {\bf 27}: 1813--1821.

\bibitem[{{\sc Smith} and {\sc Eyre-Walker}(2002)}]{Smith2002}
{\sc Smith, N. G.~C.}, and {\sc A.~Eyre-Walker}, 2002 {A}daptive protein
  evolution in {D}rosophila.
\newblock Nature {\bf 415}: 1022--1024.

\bibitem[{{\sc Strasburg} {\em et~al.\/}(2011){\sc Strasburg}, {\sc Kane}, {\sc
  Raduski}, {\sc Bonin}, {\sc Michelmore} {\em et~al.\/}}]{Strasburg2011}
{\sc Strasburg, J.~L.}, {\sc N.~C. Kane}, {\sc A.~R. Raduski}, {\sc A.~Bonin},
  {\sc R.~Michelmore}, {\em et~al.\/}, 2011 {E}ffective population size is
  positively correlated with levels of adaptive divergence among annual
  sunflowers.
\newblock Mol Biol Evol {\bf 28}: 1569--1580.

\bibitem[{{\sc Takahata} {\em et~al.\/}(1975){\sc Takahata}, {\sc Ishii} and
  {\sc Matsuda}}]{Takahata1975}
{\sc Takahata, N.}, {\sc K.~Ishii}, and {\sc H.~Matsuda}, 1975 {E}ffect of
  temporal fluctuation of selection coefficient on gene frequency in a
  population.
\newblock Proc Natl Acad Sci U S A {\bf 72}: 4541--4545.

\bibitem[{{\sc Waxman}(2011)}]{Waxman2011}
{\sc Waxman, D.}, 2011 A unified treatment of the probability of fixation when
  population size and the strength of selection change over time.
\newblock Genetics {\bf 188}: 907--913.

\bibitem[{{\sc Zhang} and {\sc Li}(2005)}]{Zhang2005}
{\sc Zhang, L.}, and {\sc W.-H. Li}, 2005 {H}uman {SNP}s reveal no evidence of
  frequent positive selection.
\newblock Mol Biol Evol {\bf 22}: 2504--2507.

\end{thebibliography}

\section*{Figures and Tables}
\clearpage
%
%

%
%

\begin{table}[h]
\caption{
\bf{$\alpha$ estimates for different fluctuating conditions with a expected mean fitness of zero. }}
\begin{center}
\begin{tabular}{lcccc}
\hline\hline
$\beta$&$\alpha_{True}$&$\alpha$ (MK$^a)$  &$\alpha$ (MK$^b$) &$\alpha$ (MK$^c$)\\ \hline
1	&0.72	&0.09	&0.22	&0.84\\
2	&0.76	&0.15	&0.26	&0.30\\
3	&0.85	&0.33	&0.39	&0.52\\
4	&0.92	&0.45	&0.52	&0.64\\
5	&0.93	&0.43	&0.50	&0.73\\
10	&0.97	&0.61	&0.68	&0.81\\
20	&0.99	&0.70	&0.74	&0.85\\
30	&0.99	&0.77	&0.81	&0.89\\
50	&1.00	&0.83	&0.86	&0.92\\
100	&1.00	&0.90	&0.93	&0.95\\
\hline\hline
\multicolumn{5}{l}{$^a$\cite{Fay2001}}\\
\multicolumn{5}{l}{$^b$\cite{Eyre-Walker2009}}\\
\multicolumn{5}{l}{$^c$\cite{Schneider2011}}\\
\end{tabular}
\end{center}
\begin{flushleft}Estimates of adaptive divergence, $\alpha$, for polymorphism and divergence simulated under varying random fluctuating selection. Three different MK type tests were used. The intensity of the fluctuation is denoted by $\beta$.
\end{flushleft}
\label{mk}
\end{table}


\end{document}